
%
%
\magnification=\magstep1

\def\dsp{\baselineskip=22pt plus 1pt minus 1pt}

\spaceskip=0.4em plus 0.15em minus 0.15em
\xspaceskip=0.5em
\hsize=16 true cm
\hoffset=0 true cm
\vsize=22 true cm

\def\sles{\lower2pt\hbox{$\buildrel {\scriptstyle <}
   \over {\scriptstyle\sim}$}}

\def\sgreat{\lower2pt\hbox{$\buildrel {\scriptstyle >}
   \over {\scriptstyle\sim}$}}

\def\lapprox{\lower2pt\hbox{$\buildrel \lower2pt\hbox{${\scriptstyle<}$}
   \over {\scriptstyle\approx}$}}

\def\gapprox{\lower2pt\hbox{$\buildrel \lower2pt\hbox{${\scriptstyle>}$}
   \over {\scriptstyle\approx}$}}

\def\both{\lower2pt \hbox{$\buildrel {\leftarrow} \over {\rightarrow}$}}

\def\etal{{\it et~al.\ }}
\def\kms{{\rm\ km~s^{-1}}} 
\def\max{{\rm max}}
\def\yr{{\rm yr}}
\def\eg{{\it e.g.\ }}
\def\b{{\rm b}}
\def\he{{\rm He}}
\def\p{{\rm p}}
\def\o{{\rm o}}
\def\cm{{\rm cm}}
\def\s{{\rm s}}
\def\recoil{{\rm recoil}}
\def\merge{{\rm merge}}
\def\obs{{\rm obs}}
\def\max{{\rm max}}
\def\yr{{\rm yr}}
\def\em{{\rm em}}
\def\ref#1{\hangindent=1cm \hangafter=1 \noindent #1}

\dsp
\centerline{\bf GAMMA-RAY BURSTS AS THE DEATH THROES OF}
\medskip
\centerline{\bf MASSIVE BINARY STARS}
\bigskip\bigskip
\centerline{Ramesh Narayan$^1$, Bohdan Paczy\'nski$^2$, and Tsvi Piran$^{1,3}$}
\bigskip
\centerline{\bf ABSTRACT}
\bigskip
It is proposed that gamma-ray bursts are created in the mergers of
double neutron star binaries and black hole neutron star binaries at
cosmological distances.  Bursts with complex profiles and relatively
long durations are the result of magnetic flares generated by the
Parker instability in a post-merger differentially-rotating disk.
Some bursts may also be produced through neutrino-antineutrino
annihilation into electrons and positrons.  In both cases, an
optically thick fireball of size $\sles\ 100$ km is initially created,
which expands ultrarelativistically to large radii before radiating.
Several previous objections to the cosmological merger model are
eliminated.  It is predicted that $\gamma$-ray bursts will be
accompanied by a burst of gravitational radiation from the
spiraling-in binary which could be detected by LIGO.
\bigskip
\ref
{{\it Subject Headings}: Accretion --- Black~Holes ---
Gamma~Rays:~Bursts --- Gravitation --- Magnetic~Fields --- Neutrinos
--- Pulsars --- Stars:~Binaries --- Stars:~Neutron ---
Stars:~Supernovae --- X-rays:~Binaries}
\bigskip
\centerline{\bf Submitted to Ap J (Lett.), March 24, 1992}
\bigskip
\ref
{$^1$ Harvard-Smithsonian Center for Astrophysics, Cambridge, MA 02138}

\ref
{$^2$ Princeton University Observatory, Princeton, NJ 08544}

\ref
{$^3$ Racah Institute for Physics, The Hebrew University, Jerusalem, Israel}

\vfill\eject
\centerline{\bf 1. INTRODUCTION}
\bigskip

Recent results obtained with the Burst and Transient Source
Experiment (BATSE) on the Compton Gamma-Ray Observatory (Meegan \etal
1992) suggest strongly that gamma-ray bursts (see Higdon \&
Lingenfelter 1990 for a review) originate at cosmological distances.
The 153 bursts reported so far appear to be isotropic in the sky and
to have a distribution of $V/V_\max$ that is consistent with a
cosmological population extending to redshifts $z\sim 1$ (Mao \&
Paczy\'nski 1992, Piran 1992).  The required event rate is $\sim
10^{-6}~\yr^{-1}$ per $L^*$ galaxy, and the typical energy released
in a burst is $\sim 10^{51}$ erg.  The short rise times
observed imply a source size $\sim 100$ km.

Based on these requirements, many authors (Paczy\'nski 1986, Goodman
1986, Eichler \etal 1989, Piran 1990, Narayan, Piran \& Shemi 1991,
Paczy\'nski 1991, Piran, Narayan \& Shemi 1992) have suggested that
$\gamma$-ray bursts arise in the merger of binaries consisting of
either two neutron stars (NS-NS) or a black hole and a neutron star
(BH-NS).  This suggestion is attractive for a number of reasons.  (1)
The model employs a known source population; four NS-NS radio pulars
have been discovered in the Galaxy, viz. PSR 1534+12, PSR 1913+16,
PSR 2127+11C, and PSR 2303+46.  (2) The scenario invokes orbital
decay through the emission of gravitational radiation, for which we
have direct observational evidence in the case of PSR1913+16 (Taylor
\& Weisberg 1989); moreover, three of the above four pulsars have
merger times shorter than the Hubble time.  (3) In a merger, an
energy $>10^{53}$ erg will be released in a time $\sim 1$ ms and
within a radius $<100$ km, satisfying the observational constraints.
(4) NS-NS and BH-NS mergers are estimated to occur at the rate of
$\sim 10^{-6}-10^{-5}~\yr^{-1}$ per galaxy (Narayan, Piran \& Shemi
1991, Phinney 1991), in good agreement with the observed burst rate.

Many arguments have been made against cosmological scenarios in
general and the merger model in particular.  Among these, one
objection appears at first sight to be quite serious.  If $10^{51}$
erg of $\gamma$-rays are created in a volume of size 100 km, the
optical depth due to $\gamma +\gamma\rightarrow e^++e^-$ will be
extremely large and the photons will apparently be trapped (Schmidt
1978).  This objection was refuted by Paczy\'nski (1986) and Goodman
(1986) who showed that an optically thick ball of energy, a
``fireball,'' will expand relativistically and thereby radiate most of
its energy when it becomes optically thin.  Because of relativistic
beaming, a distant observer receives a burst of radiation whose
temperature and duration will be similar to the initial temperature
and initial light-crossing time of the fireball.  Relativistic beaming
also circumvents the so-called ``Ruderman limit'' (Ruderman 1975),
which sets an upper limit to the distance of a source for a given
source temperature, flux, and variability timescale.  (Beaming solves
a related problem in the case of extragalactic radio jets).

Although the problems associated with the large optical depth and the
Ruderman limit have been solved, several other objections remain.
(1) How is the energy converted into $\gamma$-rays?  (2) How can one
avoid baryon contamination which will significantly modify the
evolution of the relativistic fireball?  (3) How can one obtain
bursts with a median duration $\sim 10$~s when the dynamical
timescale is $\sles ~1$ ms?  (4) Why are burst profiles so complex
and individually unique?  (5) Why do some bursts have a precursor
several seconds before the primary burst (Murakami et al.  1991)?
(6) What produces the power-law $\gamma$-ray spectrum, which extends
well beyond the pair creation limit of 511 keV and sometimes even up
to a few hundred MeV?  (7) How can one explain the cyclotron
absorption features seen in a few bursts?  (8) What is the
explanation for the redshifted electron-positron annihilation lines
claimed in some bursts?  (The evidence for these lines is not very
strong, cf. Messina \& Share 1992.)  (9) Why have no galaxies been
found in the vicinities of bright $\gamma$-ray bursts with
well-determined positions (Schaefer 1990)?

The aim of this paper is to describe a qualitative scenario for the
production of $\gamma$-ray bursts in mergers of NS-NS and BH-NS binaries,
and to demonstrate that there are plausible solutions to most of these
objections.

\bigskip\bigskip
\centerline{\bf 2. MERGER SCENARIO}
\bigskip

The progenitors of close NS-NS and BH-NS binaries must be massive
X-ray binaries consisting of an O or B main sequence star and a
neutron star or a black hole (\eg Vela X-1, Cyg X-1, cf. Trimble
1991 for a review).  When the main sequence star evolves, the binary
very likely undergoes a common envelope phase, after which one has a
tight binary consisting of the helium core of the OB star and its
compact companion.  Cyg X-3 ($P_\b$=4.79 hr) appears to be an
excellent example of this stage of evolution (van Kerkwijk \etal
1992).

We are interested in Cyg X-3 like binaries with separations $a_\o$
ranging from $\sim$ few $\times 10^{10}$ cm, the radius of the helium
core (assuming that it behaves like a helium main sequence star), to
$\sim$ few $\times 10^{11}$ cm, the limit beyond which gravitational
radiation losses in the double degenerate binary phase are too slow
to cause a merger within the age of the universe.  To make a NS-NS
binary, the helium core needs to have a mass $M_\he$ between $\sim
2.5M_\odot$ (in order to have a supernova explosion) and $4.2M_\odot$
(in order to leave behind a bound NS-NS binary, assuming a symmetric
supernova explosion and a neutron star mass of $1.4M_\odot$).  After
the explosion, we are left behind with a NS-NS binary with an orbital
eccentricity $e=(M_\he /2.8)-0.5$, periastron separation $a_\p$ equal
to the pre-explosion separation $a_\o$, and a recoil velocity
$v_\recoil=180(M_\he-2.8)(M_\he+1.4)^{-1/2}(a_\p/10^{11}\cm)^{-1/2}\kms$.
Similar estimates for a BH-NS binary give somewhat lower orbital
eccentricities and space velocities.

For typical numbers, the recoil velocity of the binary is large
enough for the system to escape from a small galaxy (though probably
not from an $L^*$ galaxy).  There is growing evidence recently that
the blue extragalactic light is dominated by faint dwarf galaxies
between $22^m - 24^m$ (Cowie 1991, and references therein).  These
galaxies are at modest redshifts ($ z \sim 0.2 - 0.4 $) and are very
numerous (1 and 6 per square arcmin at $22^m$ and $24^m$).  If
supernovae are strongly correlated with the blue light, then the
majority of compact binaries are probably born in low-mass galaxies,
escape from their hosts, and move distances of up to 1 Mpc in a few
$\times 10^9$ yr before they merge.  The lack of obvious host
galaxies associated with $\gamma$-ray bursts is then not surprising.
The burst positions are likely to be offset from that of their parent
galaxies by up to an arc minute, and at this separation the
association will be confused by several other faint galaxies in the
field.

The typical duration of a $\gamma$-ray burst is $\sim 10$ s, and the
time delay between the precursor and main pulse is also $\sim 10$ s
(Murakami et al. 1991).  For a NS-NS binary in a near-circular orbit,
the time to merge is given by $t_\merge = 38 ~ (a/10^{7.5}\cm )^4~\s =
160 ~ (P_\b /0.1\s)^{8/3}~\s$.  Although the efficiency of tidal
interactions and spin-up of the two stars as they spiral towards each
other is uncertain (Haensel, Paczy\'nski \& Amsterdamski 1991,
Kochanek 1992), it is likely that there will be sufficient heating
produced by tidal dissipation to lead to a super-Eddington luminosity.
This may produce some of the precursors.  The luminosity may also
drive a wind from the surfaces of the two stars, creating slow-moving
baryonic material which may block $\gamma$-ray emission from the
compact binary.  On the other hand, a later-produced relativistically
expanding burst may be able to shock on the slow wind to produce
non-thermal radiation.  Yet another possibility is that the precursors
may in fact be due to the early phases of the relativistic burst
material plowing through the pre-burst wind.

Once the two stars merge, there will be $\sim$ few $\times 10^{53}$
erg of energy divided more-or-less equally into three forms: (1)
thermal energy, (2) ordered rotational energy, and (3)
``non-uniform'' kinetic energy (either chaotic motions or
differential rotation).  The third form of energy may be quite
significant, and may be located quite far from the center, if there
is strong splashback during the merger as may happen if the rotations
of the two stars are not synchronized with the orbit.  Each of the
three energy forms will be released through a different channel.

The thermal energy is radiated mostly as neutrinos and antineutrinos.
Goodman \etal (1987) showed that in Type II supernovae the finite
cross-section associated with the reaction, $\nu +\bar\nu\rightarrow
e^++e^-$, results in $\sim 10^{51}$ erg being converted into
electromagnetic energy.  Eichler \etal (1989) proposed that this
mechanism may produce $\gamma$-ray bursts in binary mergers.  The
timescale of the process is the neutrino cooling time, $\sim$ few
seconds.

Although modest variability may be expected in neutrino emission
because of the chaotic (possibly convective) flows in the post-merger
fluid, the complicated and highly variable time structure observed in
some bursts may be too extreme to be explained by the neutrino
process alone.  We suggest that the neutrino mechanism produces the
sub-class of bursts with relatively smooth profiles and time
durations $\sim$ few seconds (e.g. GB830801b, Fig. 2 in Higdon \&
Lingenfelter 1990), and possibly also some precursors.

The rapid rotation of the post-merger object will almost certainly
cause the fluid to be dynamically unstable, leading to rapid loss of
energy and angular momentum on a timescale of a few msec.  This
instability dies down when the ratio of kinetic to gravitational
energy falls below $\sim 0.27$. At this stage the rotating system will
collapse directly to a black hole if the total mass is greater than
the maximum mass of a neutron star (note that some stiff equations of
state permit neutron stars more massive than $2.8_\odot$), and if the
angular momentum $J$ is smaller than $GM^2/c$.  If $Jc/GM^2>1$,
collapse is temporarily prevented, but the system will continue to
exhibit a secular instability through which it will lose angular
momentum on a timescale of a few seconds (Friedman 1983).

Most of the energy released during the first dynamical stage will be
in the form of gravitational radiation, which could be of importance
to future gravitational radiation detectors, but we do not expect any
direct electromagnetic signal.  During the second secular phase there
may be coherent emission of gravitational radiation that could be
detected by LIGO, but other processes like neutrino cooling and build
up of the magnetic field will also take place.

A fraction of the original mass will not participate directly in the
central instability and collapse, but will form a surrounding
disk/torus which may itself undergo dynamical instabilities (\eg
Narayan \& Goodman 1989).  We consider it likely that after this
phase a significant amount of residual kinetic energy, $\sles\
10^{52}$erg, will survive in a relatively long-lived
differentially-rotating disk, and we believe that this energy can be
tapped almost entirely through electromagnetic processes.

In the merger of a BH-NS binary, the neutron star will be tidally
disrupted, with some of the matter being swallowed by the black hole,
and the remainder forming a torus, generating heat and neutrinos in
the process.  Subsequent evolution may proceed in a similar fashion
in both the BH-NS and NS-NS scenarios, and it is not clear at this
time which of the two is more efficient as a $\gamma$-ray burster.

The mechanism we propose to produce $\gamma$-ray bursts similar to
GB820331 and GB841215 (Fig. 2 in Higdon \& Lingfelter 1990), which
have complex time profiles, is based on magnetic fields.  Immediately
after the merger the field strength is probably $\sim 10^{12}$ G, but
instabilities such as the Balbus \& Hawley (1991) mode will cause the
field to build up quickly to something like an equipartition value
$\sim 10^{16}-10^{17}$ G.  Even if there is no instability, the
shearing action of the differential rotation will lead to field
build-up on a slower timescale $\sim 10^4$ rotations.  (Note that,
while electric fields are limited to a maximum strength of $\sim
10^{13}$ V cm$^{-1}$ by the pair creation threshold, magnetic fields
can build up to much higher strengths.)  Once the field achieves
equipartition, it will tend to exhibit the Parker (1966) instability
in which the field will float up and break out of the disk
(relativistically in our case) on a dynamical time, leaving the matter
behind.  Particularly near the top of the magnetic loop, the baryonic
contamination may be quite low, making the conditions favorable for a
gamma-ray burst.  The burst itself will probably be created as a
series of explosive reconnection events in the rising magnetic field,
just as in solar flares.  For the assumed field strength, the
temperature of the flare will be $\sim$ few $\times 10^{11}$ K, the
photons will have characteristic energies $\sim 10$ MeV, and the total
energy in a single burst will be $\sim 10^{50}$ erg for linear
dimensions $\sim 1-10$ km.  The resulting optically thick fireball
will expand relativistically as described by Paczy\'nski (1986),
Goodman (1986), Shemi \& Piran (1990), and Paczy\'nski (1990).

A very important feature of the model is that the rate of energy
release is not limited by the Eddington luminosity, but by the
efficiency with which the Parker instability can separate magnetic
fields from matter.  The energy dissipation occurs above the disk,
and, because of the relativistic outward motion of the flare, very
little energy returns to the disk.  If this aspect of the scenario
works efficiently (which needs to be demonstrated), then we have a
viable solution to the problem of how to separate the photons from
the baryons (how to ``separate light from the darkness'').

The flare activity will make the burst profile quite complex, with
many sub-bursts.  Also, the total duration over which the successive
sub-bursts occur will be quite variable from one object to another,
depending on the details of the post-merger fluid configuration.
These features eliminate some of the strongest objections to the
merger scenario.  The important point is that the model has two
distinctly different timescales: firstly, a dynamical time, $ \sim
10^{-3} $ s, which may be related to the rise times of individual
flares (because the Parker instability, once it gets going, operates
essentially on a dynamical time), and, secondly, an accretion (or
magnetic ``viscosity'') time, $ \sim 1 - 100 $ s, related to the
duration of the whole $\gamma$-ray burst.  We cannot calculate the
second timescale from fundamental theory, but we note that in other
objects with disks, such as cataclysmic variables and X-ray binaries,
the accretion time is typically many orders of magnitude longer than
the dynamical time.

A generic problem with any cosmological model is that the huge initial
optical depth leads to themalization of the fireball (Paczy\'nski
1990) and a black body spectrum.  This is in clear conflict with the
observed spectra of $\gamma$-ray bursts, which have a significant
excess of X-rays and very hard $\gamma$-rays compared to any Planck
spectrum.  There are at least two possible solutions to this problem,
both related to the flare-like energy release and rapid variability.
First, while the instanteneous spectrum of a local flare may be
Planckian, the superposition of many Planck curves with a wide range
of temperatures and intensities may appear as a very broad broken
power law.  Second, it is likely that the flares will eject some mass
due to imperfect separation of energy and matter.  These relativistic
ejecta may collide with each other, as well as with the pre-merger
wind, at a fairly large radius and at low optical depths, giving a
non-Planckian spectrum through various non-thermal processes.  For
instance, given the strong magnetic fields, synchrotron processes
might naturally produce the observed power-law spectrum if enough
relativistic particles are present.

An important point to note is that the observed photon energy
$h\nu_\obs$ is related to the emitted energy in the frame of the
radiating fireball $h\nu_\em$ by the large relativistic $\gamma \sim
10^2-10^3$ of the expansion (Paczy\'nski 1986, Goodman 1986).
Therefore, there is no difficulty in producing very hard spectra.  In
particular, note that a cut-off at 511 keV in the emitter frame
translates to a cut-off at $\gamma \times 511$ keV for the observer,
which can correspond to several hundred MeV.

The fact that the energy release in the model is predominantly
through a magnetic process may solve another potential problem, the
cyclotron absorption lines that have been claimed in some bursts.  In
very rough terms, an initial field of $10^{16}$ G at the point of
initiation of the burst will be diluted to a field strength $\sim
10^{10}$ G in the frame of the emitting fluid if the expansion factor
is $\sim 10^3$.  However, because of relativistic beaming, the
observed cyclotron line will appear as if it is emitted in a field
$\sim \gamma \times 10^{10} $ G $\sim 10^{13}$ G, which is roughly
consistent with the claimed field strengths (Murakami et al.  1988,
Fenimore et al. 1988).  The narrowness of the observed lines requires
a stable value of the Lorentz $ \gamma $ for the duration of
the line.  This will be difficult to maintain for any length of time
in our highly erratic burster, in qualitative accordance with the
``now you see them, now you don't'' nature of the reported lines.

Another interesting possibility should be finally mentioned.  As the
disk cools through neutrino emission, it might at some point make a
transition to a superconducting superfluid state.  Plasma processes in
a degenerate superconducting superfluid have been hardly studied at
all.  It is conceivable that the separation of the field and the
baryons may be particularly efficient in a magnetic superfluid.  This
problem merits further investigation.

\bigskip\bigskip
\centerline{\bf 3. CONCLUSIONS}
\bigskip

We propose a scenario for gamma-ray bursts involving merging NS-NS
and BH-NS binaries at cosmological distances.  In our picture,
complex main bursts with highly variable time profiles are produced
through many successive magnetic flares and reconnection events
tapping the free energy in differential rotation in a post-merger disk.
Less variable bursts may be produced by neutrinos converting a
fraction of their energy into electromagnetic radiation.  The
precursors of the bursts may be produced by the neutrino mechanism,
or by the initial phases of the post-merger ultra-relativistic burst
ejecta plowing through the slow wind created in the pre-merger tidal
heating phase, or even possibly by the pre-merger heating itself.
This scenario is capable of explaining the qualitative features of
most of the observations, and circumvents many of the arguments
against the cosmological merger model of gamma-ray bursts.

The model employs two kinds of objects, NS-NS and BH-NS binaries, and
two distinct mechanisms, magnetic flares and neutrino interactions.
Different combinations of these may have distinct observational
signatures which could be potentially identified in the large
database of burst profiles being acquired by BATSE.

Apart from $\gamma$-rays, it is likely that the bursts in our scenario
will also eject $ \sim 10^{51} $ erg of energy in
ultra-relativistic ejecta, i.e. cosmic rays, with little or no
non-relativistic matter.  These ejecta should create a supernova-like
remnant far from the parent galaxy, where the density of diffuse
matter may be very low.

The scenario makes a definite prediction.  If and when LIGO is
commissioned (Vogt 1992), strong $\gamma$-ray bursts should be
accompanied by gravity wave detections (though the reverse need not
necessarily be true, particularly if the $\gamma$-rays are beamed).
Since the strength of the gravitational radiation signal from a
pre-merger binary can be calculated accurately (Lincoln \& Will 1990)
except for the unknown inclination, LIGO should provide good distance
estimates to individual bursts and also pinpoint the exact time of the
merger.  This information will permit more detailed interpretation of
the $\gamma$-ray burst observations.

The scenario we have presented in this paper is at this point no more
than a qualitative sketch of some likely possibilities.  The purpose
here is merely to show that none of the many arguments against the
merger model are necessarily fatal.  More detailed analyses of the
individual ideas are needed now.

One other point worth emphasizing is that the scenario is in a real
sense the most conservative one possible at the present time.  Based
purely on the observations, a cosmological location for $\gamma$-ray
bursts is the most reasonable hypothesis; all other proposals need
coincidences at some level in the properties of the associated source
population.  Within the cosmological framework, our scenario employs
the most obvious source population that we can think of, double
compact binaries.  This population is definitely known to exist, we
know its members will merge, and we can be certain that huge
quantities of energy will be released.  However, even if this
particular model turns out to be incorrect, the cosmological
hypothesis itself will survive, given the observed distribution of
burst positions and intensities.

We thank Roger Blandford and John Friedman for discussions and
comments.  This work was supported by NASA grants NAG 5-1901 and NAG
5-1904.

\vfill\eject
\centerline{\bf REFERENCES}
\bigskip
\ref
{Balbus, S. A. \& Hawley, J. F. 1991, ApJ, 376, 214}

\ref
{Cowie, L. L. 1991, in Primordial Nucleosynthesis and Evolution of Early
    Universe, Astrophysics \& Science Library, Vol. 169, p. 425 (Eds.:
    K. Sato \& J. Audouze, Kluwer Academic Publishers)}

\ref
{Eichler, D., Livio, M., Piran, T. \& Schramm, D. N. 1989, Nature, 340, 126}

\ref
{Fenimore, E. E. et al. 1988, ApJ, 335, L71}

\ref
{Friedman, J. 1983, Phys. Rev. Lett., 51, 11}

\ref
{Goodman, J. 1986, ApJ, 308, L47}

\ref
{Goodman, J., Dar, A., \& Nussinov, S.  1987, Ap. J. Letters, 314, L7}

\ref
{Haensel, P., Paczy\'nski, B., \& Amsterdamski, P. 1991, ApJ, 375, 209}

\ref
{Higdon, J. C. \& Lingenfelter, R. E. 1990, ARAA, 28, 401}

\ref
{Kochanek, C. S. 1992, ApJ, submitted}

\ref
{Lincoln, C. W. \& Will, C. M. 1990, Phys. Rev., D42, 1123}

\ref
{Mao, S. \& Paczynski, B. 1992, ApJ (Lett.), in press}

\ref
{Meegan, C. A. \etal 1992, Nature, 355, 143}

\ref
{Messina, D. C. \& Share, G. H. 1992, Proc. Huntsville GRO
Meeting, eds. W. S. Paciesas \& G. J. Fishman, in press}

\ref
{Murakami, T., et al. 1988, Nature, 335, 234}

\ref
{Murakami, T., et al. 1991, Nature, 350, 592}

\ref
{Narayan, R. \& Goodman, J. 1989, Proc. NATO Advanced Research Workshop
on ``Theory of Accretion Disks'', eds. F. Meyer, W. J. Duschl, J.
Frank \& E. Meyer-Hofmeister, p. 231}

\ref
{Narayan, R., Piran, T. \& Shemi, A. 1991, ApJ, 379, L17}

\ref
{Paczy\'nski, B. 1986, ApJ, 308, L51}

\ref
{Paczy\'nski, B. 1990, ApJ, 363, 218}

\ref
{Paczy\'nski, B. 1991, Acta Astron., 41, 257}

\ref
{Parker, E. N. 1966, ApJ, 145, 811}

\ref
{Phinney, E. S. 1991, ApJ, 380, L17}

\ref
{Piran, T. 1990, in Supernovae, eds. J. C. Wheeler, T. Piran \&
S. Weinberg, World Sci. Publ., Singapore}

\ref
{Piran, T. 1992, ApJ (Lett.), in press}

\ref
{Piran, T., Narayan, R. \& Shemi, A. 1992, Proc. Huntsville GRO
Meeting, eds. W. S. Paciesas \& G. J. Fishman, in press}

\ref
{Ruderman, M. A. 1975, Ann. N Y Acad. Sci., 262, 164}

\ref
{Schaefer, B. E. 1990, Los Alamos Workshop on Gamma-Ray Bursts, July 29 -
  Aug. 3, 1990}

\ref
{Schmidt, W. K. H. 1978, Nature, 271, 525}

\ref
{Shemi, A. \& Piran, T. 1990, ApJ, 365, L55}

\ref
{Taylor, J. H. \& Weisberg, R. M. 1989, ApJ, 345, 434}

\ref
{Trimble, V. 1991, Cont. Phys., 32, 103}

\ref
{van Kerkwijk, M. H. \etal 1992, Nature, 355, 703}

\ref
{Vogt, R. 1992, in Proc. MG6, eds. K. Sata \& T. Nakamura, World
Sci. Publ., Singapore, in press}

\bye